\documentclass[a4paper]{article}

\usepackage{format}
\usepackage[english]{babel}

\title{DAMIC: a novel dark matter experiment}

\shorttitle{DAMIC: Dark Matter In CCDs}

\authors{
Javier Tiffenberg$^{1}$
for the DAMIC Collaboration.
}

\afiliations{
$^1$ Fermi National Accelerator Laboratory, Batavia, Illinois 60510, USA.
}

\email{javiert@fnal.gov}

\abstract{DAMIC (Dark Matter in CCDs) is a novel dark matter experiment that has unique sensitivity to dark matter particles with masses below 10~GeV. Due to its low electronic readout noise (R.M.S.~$\sim$3~e-) this instrument is able to reach a detection threshold below 0.5~keV nuclear recoil energy, making the search for dark matter particles with low masses possible. We report on early results and experience gained from a detector that has been running at SNOLAB from Dec 2012. We also discuss the measured and expected backgrounds and present the plan for future detectors to be installed in 2014.}


\begin{document}
\maketitle

\section{Introduction}
The evidence for Dark Matter (DM) has been well established by astronomical observations. This has
produced a large experimental program to directly detect dark matter in the laboratory. Most of these
experiments have been optimized for Weakly Interactive Massive Particles (WIMPs) with mass above 50~GeV.
This large mass is motivated by the minimal supersymmetric extensions to the standard model.
Detection thresholds of a few keV nuclear recoil energy (keV$_r$)
are typical for such high mass DM searches, and these experiments are
very close to excluding the most natural region of parameter space consistent with a supersymmetric
particle. Various other models relating dark matter with the baryon asymmetry favor light dark matter masses $\sim$5~GeV~\cite{bib:LDM}.

The DAMIC experiment is a DM search using CCDs that can operate at a threshold below 0.5~keV$_{r}$.
The DAMIC team has performed an engineering run to demonstrate the technology in a shallow
underground site at Fermi National Accelerator Laboratory using a detector with 0.5~g of active mass. The
results of this run at the shallow site have produced the best limits for DM searches with mass below 4~GeV~\cite{bib:DAMIC_MINOS}. We are now conducting a new run at SNOLAB with an active mass
of 5~g and lower background. 

We are also engaged in building the next version of the experiment (DAMIC-100) that will have sensitivity to
cover the region of the DM parameter space consistent with the recent experimental hints for DM and
achieve the world best sensitivity for masses below 10~GeV, opening up a region in the parameter space
inaccessible to other experiments.

\subsection{Status of Direct DM searches and possible hint for low mass dark matter}
\begin{figure}[t]
  \centering
  \includegraphics[width=0.48\textwidth]{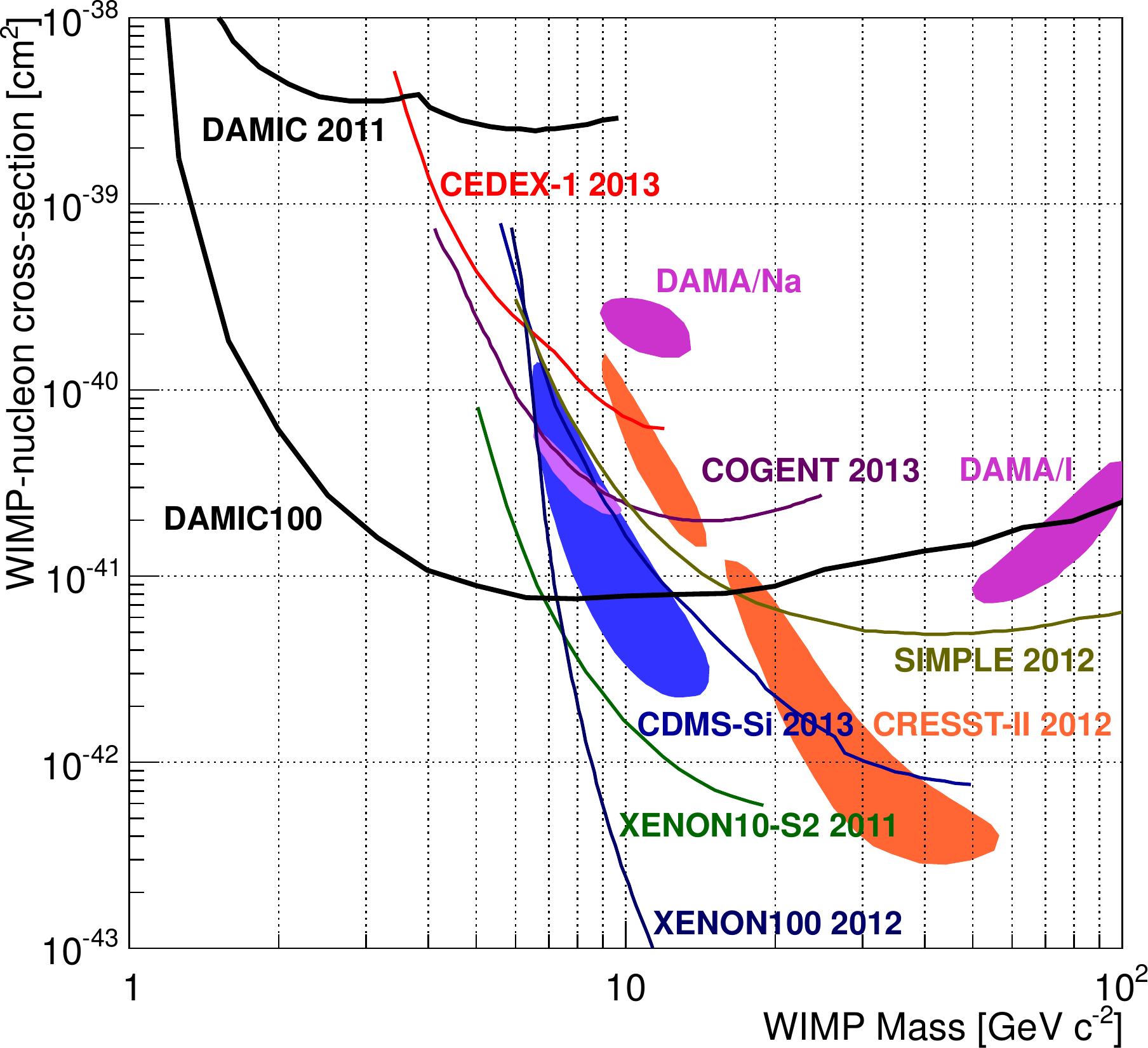}
  \caption{Current 90\%~C.L. exclusion limits and regions consistent with observed excess of events for spin independent
dark matter searches by several Collaborations~\cite{bib:DAMIC_MINOS, bib:CDMS_Si, bib:COGENT, bib:CRESSTII, bib:DAMA, bib:XENON, bib:CEDEX}. Also shown is the expected sensitivity range for the proposed DAMIC100 after 1 year of data taking.}
  \label{img:limits}
 \end{figure}

The current status of direct DM searches in the low mass range is summarized in figure~\ref{img:limits}.
The figure shows that there is a set of experiments that have produced limits
that are probing into the high mass region preferred by supersymmetric model that quickly
lose sensitivity as the DM particle mass is reduced. At the same time there are claims of
experimental results consistent with the observation of a DM signal~\cite{bib:CDMS_Si, bib:COGENT, bib:CRESSTII, bib:DAMA}. 
There is significant tension between these results, because the region of positive signals is rejected to high confidence level by some experiments. 
In particular, the XENON-100 experiment claims rejection to 90\%~C.L. of all the experimental
hints for DM~\cite{bib:XENON}. There has been extended discussion about these results in the literature, including speculation of possible systematic effects that could influence the different experimental results~\cite{bib:14, bib:15, bib:16}.

In summary, for more than 10 years the DM community has been puzzled by a possible detection
of a low mass DM particle by DAMA/NaI and DAMA/LIBRA that could not be confirmed experimentally.
Since the supposed signal was detected using lower energy thresholds than 
was used in the the other experiments which excluded their results, their signal has not been 
completely ruled out as Dark Matter.
In the past two years there have been three new experimental hints for the low mass dark matter signal as shown in figure~\ref{img:limits}.

\section{Charge coupled devices as a new tool in the search for low mass dark matter}
Operating at temperatures below 175~K, CCDs used for astronomical imaging and spectroscopy 
commonly achieve readout noise levels of 2~e-~R.M.S., 7.3~eV of ionizing energy 
in silicon (electron equivalent energy or eVee).
Although this allows a very low detection threshold, these detectors have not been 
previously considered for DM search because of their low active mass.
The development of thick fully-depleted CCDs, ten times more massive than conventional
CCDs, has changed this situation. The DAMIC experiment is the first DM
experiment to exploit this technology.

The microdetector group at Lawrence Berkeley National Laboratory (LBNL) has developed
detectors with a depletion region 250~$\mu$m thick fabricated with high resistivity silicon~\cite{bib:20}.
These devices have been used to build the focal plane of the Dark Energy Camera
(DECam)~\cite{bib:21, bib:22} and other astronomical instruments~\cite{bib:23, bib:24, bib:25}.
The idea of using them in a DM experiment was motivated by the early R\&D work with DECam. The two main features that make the
DECam CCDs good candidates for a direct dark matter search are their thickness, which allows the CCD
to have significant mass, and the low electronic readout noise, which permits a very low energy threshold
for the ionization signal produced by nuclear recoils.

\subsection{Calibration and response to Nuclear Recoils}
\mbox{X-rays} are commonly used for the energy calibration of CCD detectors~\cite{bib:26}. \mbox{X-rays} from $^{55}$Fe
penetrate only $\sim$20~$\mu$m into the silicon before producing charge pairs according to the well known
conversion factor of 3.64~eVee/e- for ionization in Si~\cite{bib:26}. As a result of this process \mbox{X-rays} produce hits in
the detector with a size limited by charge diffusion. In a DECam CCD, the size of the hit produced by 5.9~keV \mbox{X-rays} that interact in the back of the CCD (maximum diffusion) is $\sim$7~$\mu$m~R.M.S.~\cite{bib:27}.

\begin{figure}[t]
  \centering
  \includegraphics[width=0.48\textwidth]{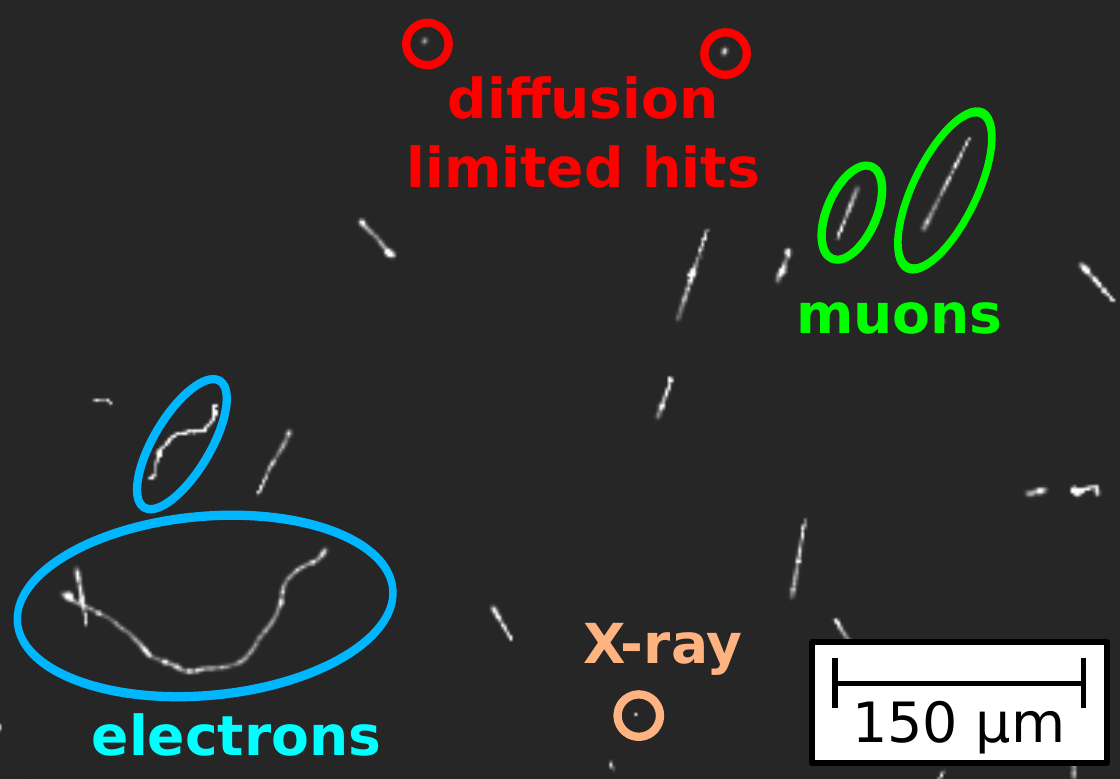}
  \caption{Sample of tracks recorded during a short exposure at sea level. The tracks produced by different kinds of particles are easily identified.}
  \label{img:tracks}
 \end{figure}
 
\begin{figure}[t]
  \centering
  \includegraphics[width=0.48\textwidth]{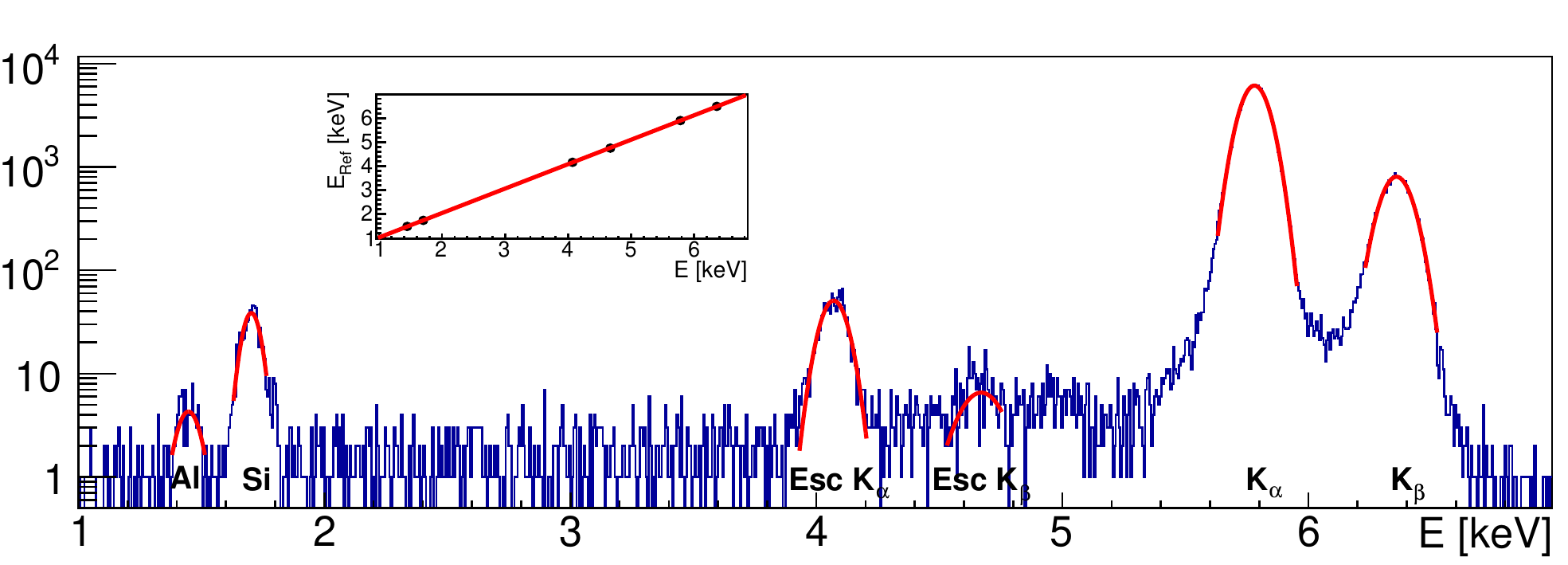}
  \caption{Energy spectrum obtained from reconstructed \mbox{X-ray} hits using an $^{55}$Fe source. The K$_{\alpha}$ and K$_{\beta}$ peaks of the Mn (5.9~keV and 6.5~keV) and their corresponding escape peaks (4.16~keV and 4.75~keV) can be used for calibration. Also useful are the Si (1.74~keV) and Al (1.48~keV) fluorescence peaks. For this calibration run the CCD was contained in an aluminum box.}
  \label{img:xray}
 \end{figure}

The total charge and shape of each hit is extracted using dedicated image analysis tools.
The \mbox{X-ray} hits are easily identified in the CCD images (fig.~\ref{img:tracks}). The energy spectrum measured for an $^{55}$Fe \mbox{X-ray} exposure 
is shown in fig.~\ref{img:xray}.

Because dark matter particles are expected to be electrically neutral and have low kinetic energy, their main interaction channel is with the Si nuclei through coherent interaction and not with the electrons. 
On the other hand, the \mbox{X-ray} photons interact mainly with the electrons around the Si
nuclei. Figure~\ref{img:xray} provides a good calibration for electron recoils in Si, produced by \mbox{X-rays}, but it is
not directly applicable to nuclear recoils. The ratio between the ionization efficiency for nuclear recoils
and electron recoils is usually referred to as the quenching factor $Q$, and has been measured in Si for
recoil energies above 4~keV$_r$~\cite{bib:quenching}. The results show good agreement with the Lindhard 
model~\cite{bib:31, bib:32}, but there is currently no data for recoil energies below 4~keV$_r$. At these low energies the quenching factor becomes increasingly energy-dependent.
We have several ongoing efforts to measure the quenching factor at energies below 4~keV$_r$.

\section{Readout Noise on High Resistivity CCDs}
The noise performance of these detectors has been studied in detail as part of the characterization
effort done by the DECam team~\cite{bib:29}. Two amplifiers in parallel read out the detectors. The amplifiers are located
on opposite ends of a serial register toward which the charge is clocked. The detectors have an
output stage with an electronic gain of \mbox{$\sim$2.5~$\mu$V/e-.} The signal is digitized after correlated double
sampling (CDS) of the output. The CDS removes the noise resulting from resetting the output stage after
each pixel, and is also efficient in rejecting the common mode noise (frequencies much lower than the pixel
frequency). Each sample used for the CDS operation is the result of integration during a time window. The
integration time can be selected freely and acts as a filter for high frequency noise. The
noise observed for pixel readout times larger than 30~$\mu$s is $<2$~e- (RMS). These results were obtained
using a Monsoon~\cite{bib:29} CCD controller. 
This performance is common for scientific CCDs, but exceptional for a
DM detector. Having a readout noise of 2~e- RMS corresponds to a noise of 7.3~eVee.
Such a low noise makes it possible to operate a DM experiment
with a threshold below 0.5~keV$_{r}$. This is the main reason to consider CCDs for a DM search.

At the present, we are operating the setup currently running at SNOLAB at a higher threshold of $\sim$100~eVee due to the lack of electrical insulation between the vessel and the cryo-cooler. We are currently building custom made connectors that will isolate the detector from the refrigerator. The new connectors will be installed during August 2013.

\begin{figure*}[t!]
  \centering
  \includegraphics[width=\textwidth]{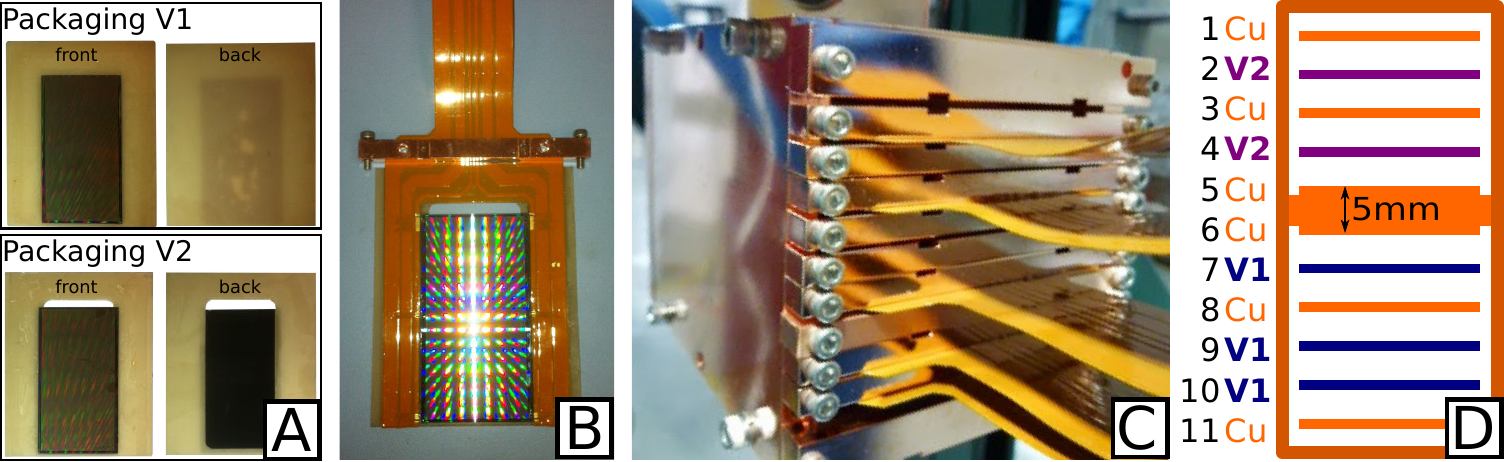}
  \caption{A)~Version 1~(top) and 2~(bottom) of the low background package developed for the SNOLAB tests. In the later version the AlN mass is greatly reduced by removing most of the substrate material from the active area of the CCD keeping only a AlN frame.
  B)~Complete package with long flex circuit that allows placing the connectors outside the shield.
  C)~Stack of 5~CCD detectors inside the copper box ready for installation into the vacuum vessel for operation at SNOLAB.
  D)~CCD detectors ordering inside the copper box. The CCD in slot 10 is facing up, it has AlN substrate below it and on top of it (from the CCD in slot 9). The two CCDs using package V2 are enclosed between copper slabs. }
  \label{img:pack}
\end{figure*}

\section{Running at SNOLAB}
The SNOLAB underground laboratory in Canada has low intrinsic background due to its 6000~m.w.e.
overburden. At this level we expect a rate of 0.26~m$^{-2}$~day$^{-1}$ cosmic muons~\cite{bib:SNOLABBack} which
strongly suppresses the cosmogenic neutrons that constituted our main source of irreducible background in 
our previous experiment in the NuMI near detector hall at FNAL which is only at a 350~m.w.e. depth.
The neutron flux at SNOLAB has been measured to be 4000~n~m$^{-2}$~day$^{-1}$ with a mean energy of 2.5~MeV~\cite{bib:SNOLABBack}.

Simulations using MCNPX and GEANT4 have been developed to estimate the background rate expected at SNOLAB. After consideration of the radioactive contamination of all materials in the shield and backgrounds from the SNOLAB environment, we expect a contribution below 1~cpd~kg$^{-1}$~keVee$^{-1}$ from materials external to the inner copper box. At the NuMI detector hall in FNAL, the DAMIC prototype observed 600~cpd~kg$^{-1}$~keVee$^{-1}$ below 2~keVee. We have observed an order of magnitude background reduction during our current run at SNOLAB, mostly due to the redesigned CCD package (Section~\ref{sec:package}). The ongoing tests at SNOLAB, together with a radioactive screening program of the packaging materials, should allow us to reach background levels below 1~cpd~kg$^{-1}$~keVee$^{-1}$.

\subsection{CCD package}
\label{sec:package}
For the run at SNOLAB a new low background detector package was developed. The main improvement 
of this package over the version used at the NuMI engineering run is that the connector was moved to be outside the lead shield, reducing the material with possible active isotopes close to the sensor.
At the same time the new package allows for easily stacking a set of detectors to achieve a higher mass. This stack of five detectors inside the copper box is shown in fig.~\ref{img:pack}. The long flex cables allow the electronic connectors to be outside the lead shield (fig.~\ref{img:shield}).

\begin{figure}[t]
  \centering
  \includegraphics[width=0.48\textwidth]{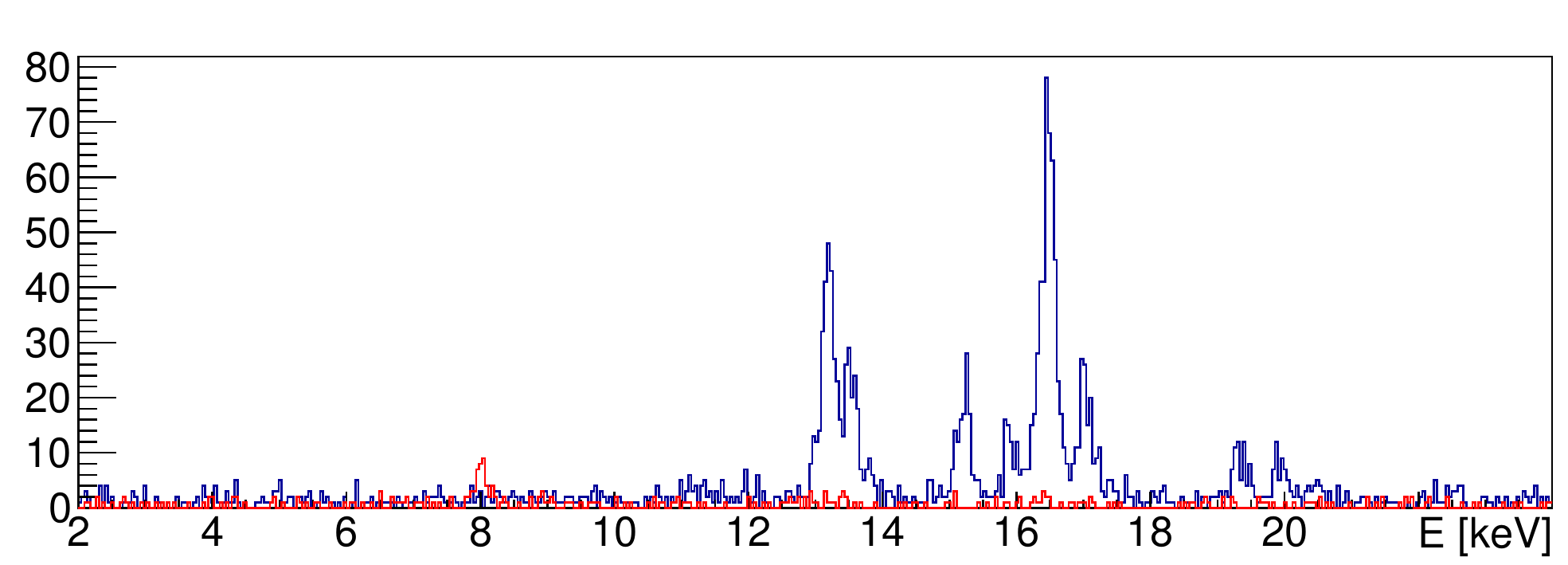}
  \caption{Raw energy spectra (no selection cuts) for an exposure of 10~g~day$^{-1}$ recorded by the CCDs in slot 10 using package~V1 (blue) and in slot 2 using package~V2 (red). The peaks in the spectrum from CCD in slot 10 match the expectation for the \mbox{X-rays} produced by uranium contamination in the AlN substrate. The second version of the package greatly reduce the effect of the contamination in the AlN. The Cu fluorescence peak (8~keVee) is visible in the spectrum recorded by the CCD in slot 2 because its active area is directly exposed to copper (fig.~\ref{img:pack}~D).}
  \label{img:spectra}
 \end{figure}

The first version of the new package (V1) consisted of a rectangular aluminum nitride (AlN) substrate 
that provides mechanical support for the CCD and a custom made U-shaped flex cable (fig.~\ref{img:pack}).
During a preliminary run in January 2013 we found that the AlN substrate had traces of $^{238}$U with an activity $\sim$3~Bq~kg$^{-1}$ (fig.~\ref{img:spectra}).
This background was the limiting factor for the detector sensitivity and prevented 
the identification of other potential contaminations in the materials that are close to the 
CCD detectors. For this reason we developed a second version of the package (V2) in which the AlN mass was reduced by removing most of the substrate material from the active area of the CCD keeping only an AlN frame (fig.~\ref{img:pack}). This second version of the package greatly reduced the effect of the contamination in the AlN (fig.~\ref{img:spectra}).
For future versions we plan to replace the AlN by high purity silicon and reduce 
the mass of the flex cable.

\subsection{Shield}
\begin{figure}[t]
  \centering
  \includegraphics[width=0.48\textwidth]{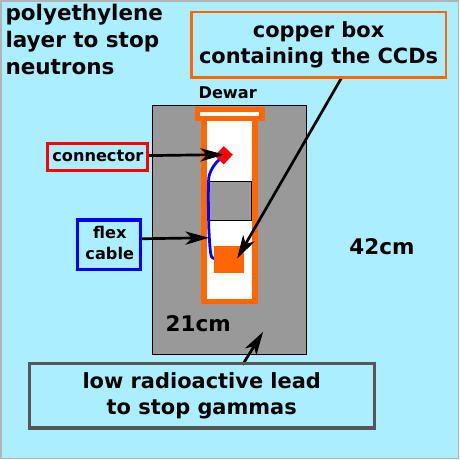}
  \caption{schematic view of the detector shield.}
  \label{img:shield}
\end{figure}

Figure~\ref{img:shield} shows a schematic view of the detector shield.
The CCDs are contained in a high purity copper box 5~mm thick cooled to 145~K, the operation temperature of the CCDs. This innermost layer of material blocks the low energy radiation and shields the CCDs from thermal photons produced in the dewar vessel wall which is at room temperature.
The vessel is surrounded by a 21~cm thick lead wall to reduce the gamma radiation. The lead was extracted from the Doe Run mine which has lower natural radioactivity than commercial lead~\cite{bib:doerun}. The lead shield is surrounded by a 42~cm thick polyethylene layer to stop neutrons coming from the cavern walls. Our simulations show that the shield reduces the outside radiation below 1~cpd~kg$^{-1}$~keVee$^{-1}$. The remaining background radiation is produced by materials inside the detector vessel.

\section{DAMIC-100}
We are currently working on the design and construction of a CCD dark matter experiment with 100~g of active
mass that will be installed at SNOLAB during 2014.

The LNBL microdectors group has recently produced fully depleted 500~$\mu$m CCDs with 6~cm~x~6~cm active
area that are the baseline detectors for the BigBOSS spectrograph~\cite{bib:26}. These sensors have 4 times the
active mass of the DECam CCDs. The proposed DAMIC-100 would use 25 of these sensors to
achieve 100~g active mass.

Most of the effort during the current run at SNOLAB is focused on understanding the activity of the inner component materials of the detector to design a package that would allow us to get full advantage of the mass increase. Figure~\ref{img:limits} shows the expected sensitivity of DAMIC-100  after a 1~year run (35.6~kg~day$^{-1}$).

\end{document}